\documentclass[aps,prl,twocolumn,superscriptaddress]{revtex4-1}

\usepackage{amsmath}
\usepackage{graphicx}
\usepackage{dsfont}
\usepackage{changepage}
\usepackage{fancyhdr}
\usepackage{amsthm, amssymb}
\usepackage{array}
\usepackage[usenames,dvipsnames]{color}

\newcommand{\be}{\begin{align}}
\newcommand{\ee}{\end{align}}

\def \be{\begin{equation}}
\def \ee{\end{equation}}
\def \ba{\begin{array}}
\def \ea{\end{array}}
\def \bea{\begin{eqnarray}}
\def \eea{\end{eqnarray}}

\def \a{{\alpha}}

\def \ba{\begin{align*}}
\def \ea{\end{align*}}

\newcounter{indice}

\def \mc{\mathcal}

\begin{document}


\title{Chiral superconductivity in the alternate stacking compound 4Hb-TaS$_2$}


\author{A. Ribak}
\affiliation{Physics Department, Technion-Israel Institute of Technology, Haifa 32000, Israel}
\author{R. Majlin Skiff}
\affiliation{Raymond and Beverly Sackler School of Physics and Astronomy, Tel-Aviv University, Tel Aviv, 69978, Israel}
\author{M. Mograbi}
\affiliation{Raymond and Beverly Sackler School of Physics and Astronomy, Tel-Aviv University, Tel Aviv, 69978, Israel}
\author{P.K. Rout}
\affiliation{Raymond and Beverly Sackler School of Physics and Astronomy, Tel-Aviv University, Tel Aviv, 69978, Israel}
\author{M. H. Fischer}
\affiliation{Department of Physics, University of Zurich, 8057 Zurich, Switzerland}
\author{J. Ruhman}
\affiliation{Department of Physics, Bar-Ilan University, Ramat Gan 5290002, Israel}
\author{K. Chashka}
\affiliation{Physics Department, Technion-Israel Institute of Technology, Haifa 32000, Israel}
\author{Y. Dagan}
\affiliation{Raymond and Beverly Sackler School of Physics and Astronomy, Tel-Aviv University, Tel Aviv, 69978, Israel}
\author{A. Kanigel}
\affiliation{Physics Department, Technion-Israel Institute of Technology, Haifa 32000, Israel}


\date{\today}
%


\maketitle


\textbf{Layered van der Waals (vdW) materials are emerging as one of the most {versatile directions} in the field of quantum condensed matter physics.  They allow an unprecedented control of electronic properties via stacking of {different} types of two-dimensional (2D) {materials}~\cite{Geim_2D, novoselov:2016} and, moreover, by tuning the relative angle between them\cite{Pablo1,Pablo2}. A fascinating frontier, {largely} unexplored, is the stacking of strongly-correlated phases of matter in vdW materials.
Here, we study 4Hb-TaS$_2$, which {naturally} realizes an alternating stacking of a Mott insulator, recently reported as a gapless spin-liquid candidate(1T-TaS$_2$)\cite{Lee,Ribak2017}, and a 2D superconductor competing with charge-density wave order (1H-TaS$_2$)\cite{yang2018enhanced}. {This raises the question of how these two components affect each other.}
We find a superconducting ground state with a transition temperature of 2.7K, which is significantly elevated compared to the 2H polytype (Tc=0.7K). Strikingly, the superconducting state exhibits signatures of time-reversal-symmetry breaking abruptly appearing at the superconducting transition, which can be naturally explained by a chiral superconducting state.}

{ Chiral superconductors (SCs) have received much attention in recent years as a promising platform for hosting Majorana bound states in the vortex cores or at sample edges due to the topological nature of their ground states~\cite{ReadGreen,sau2010generic}. In 2D, chiral superconductors are characterized by a Chern number. The Majorana bound states are predicted to possess non-Abelian statistics, which makes them candidates for performing fault-tolerant quantum computations~\cite{NayakReview}.
The order parameter of these chiral states break time-reversal symmetry (TRS), which manifests itself at edges and defects~\cite{Ueda} and can be detected with probes such as muon spin relaxation and polar Kerr effect.}

Of all the known {superconductors}, only few exhibit signatures of TRS breaking, and even fewer {are candidates} for this elusive chiral phase. {The best known among them are the spin-triplet superconductors Sr$_2$RuO$_4$, believed to be of $p+ip$ symmetry~\cite{Luke_SRO} and UPt$_3$~\cite{schemm2014observation}, a potential $f+if$ superconductor, as well as the spin-singlet heavy-fermion superconductors URu$_2$Si$_2$~\cite{kawasaki2014} and SrPtAs ~\cite{SrPtAs_muSR,Fischer2014}, which were suggested to be of $d+id$ symmetry.  Open questions remain, however, in all cases~\cite{Hassinger2017,Schemm2017}.}

In this work, we show evidence for chiral superconductivity in the transition-metal dichalcogenide (TMD) 4Hb-TaS$_2$.
We show that this polymorph of TaS$_2$ is a superconductor with a relatively high T$_c$, anomalous transport properties and that it exhibits a spontaneous appearance of  magnetic moments with the onset of superconductivity.

\begin{figure}[h!]
	\centering	
	\includegraphics[width=80mm]{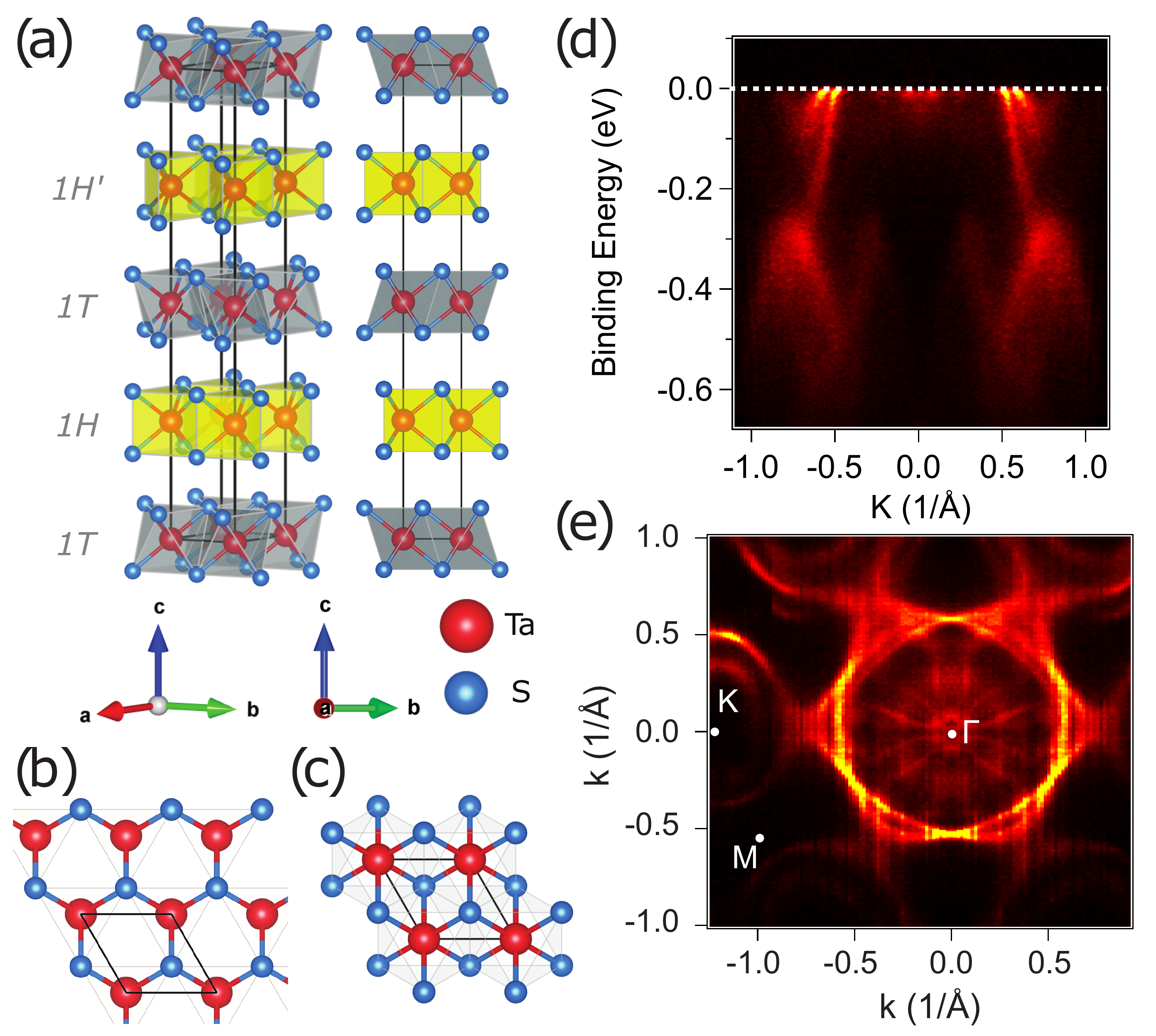}
	\caption{(a) Three-dimensional schematic drawing of a unit cell of 4Hb-TaS$_2$ showing the alternate stacking of octahedral (T) and trigonal prismatic (H) layers. A top view of 1H and 1T layers are shown in (b) and (c) respectively. The top view of the 1H layer displays the in-plane  broken mirror symmetry.(d) An ARPES detector image obtained at T=15 K using 72 eV photon energy, reveals the electronic band structure along the $\Gamma-M$ direction.
(e)  A Fermi surface mapping at the same conditions. The band structure is a combination of 2H-TaS$_2$ and CDW reconstructed 1T-TaS$_2$ which was rigidly shifted towards the Fermi level (horizontal white dashed line in (d).}
	\label{Struct}
\end{figure}

4Hb-TaS$_2$ belongs to { the} $P63/mmc$ hexagonal space group, with a unit cell that consists of alternating layers of 1H-TaS$_2$ (half of 2H-TaS$_2$) and 1T-TaS$_2$, see Fig.~\ref{Struct}(a).
The overall crystal is inversion symmetric, with the inversion point lying in the center of the 1T layer. The weak interlayer coupling allows to describe 4Hb-TaS$_2$ as a stack of 2D monolayers: 1H-TaS$_S$ with a locally broken inversion symmetry giving rise to antisymmetric spin-orbit coupling (ASOC) and 1T-TaS$_2$ that is known as a Mott insulator that fails to order magnetically \cite{Fazekas1980}. Recently, it was proposed that the ground state of 1T-TaS$_2$ is a gapless quantum spin liquid \cite{Lee,Ribak2017,Matsuda_thermal}. Thus, 4Hb-TaS$_2$ is { a} system in which superconducting layers {naturally} sit in proximity to layers that { possess} strong spin-fluctuations. Consequently, it realizes a unique heterostructure of strongly correlated phases with drastically different ground states albeit having the exact same chemical composition and almost the same structure.

4Hb-TaS$_2$ was first synthesized by Di Salvo \textit{et al} \cite{DiSalvo_4Hb}.
The transport data can be described as a mixture of 1T and 2H, with metallic conductivity in the $ab$ plane and semiconducting conductivity along the $c$ axis. The in-plane resistivity was shown to be three orders of magnitude smaller than the out-of-plane resistivity. This ``mixture'' of 1T and 2H is also visible in the X-ray photoelectron spectroscopy spectrum of the Ta $4f$ core-levels, which displays three peaks - two from the `1T' layer and one from the `1H' layer \cite{TaS2-1T_4H-XPS}. { (Supplementary Fig. S2)}

We have grown single crystals of 4Hb-TaS$_{2-x}$Se$_{x}$ with about 1\% of Se, using a standard chemical vapor transport method \cite{DiSalvo_4Hb}. The small amount of Se stabilizes the 4Hb structure. Details about the sample preparation process and the structure characterization are given in the { supplementary} material.

The electronic dispersion along the $\Gamma-M$ direction was measured using angle-resolved photoemission spectroscopy (ARPES), and is presented in Fig~\ref{Struct}(d). The band structure is a mixture of the band structure of  2H-TaS$_2$, with its two electron pockets centered at $\sim\textbf{k}=\pm0.77\AA^{-1}$,
and of the band structure of 1T-TaS$_2$. Interestingly, the 1T part of the band structure is reconstructed by the well known $\sqrt{13} \times \sqrt{13}$ charge density wave (CDW) but it is shifted towards the Fermi level leaving no spectral gap.

An ARPES intensity map at $E_F$ is shown in Fig.~\ref{Struct}(e)	. The intensity map reveals that within the alternate-stacking layered crystal, every layer retains its original electronic dispersion. The Fermi surface is  a mixture of the three-fold symmetric Fermi surface of 2H-TaS$_2$ with its familiar dog-bone shaped pockets around the $M$ points, and the six-fold reconstructed 1T-TaS$_2$ bands which were shifted towards the Fermi level.

\begin{figure}[h!]
	\centering	
	\includegraphics[width=80mm]{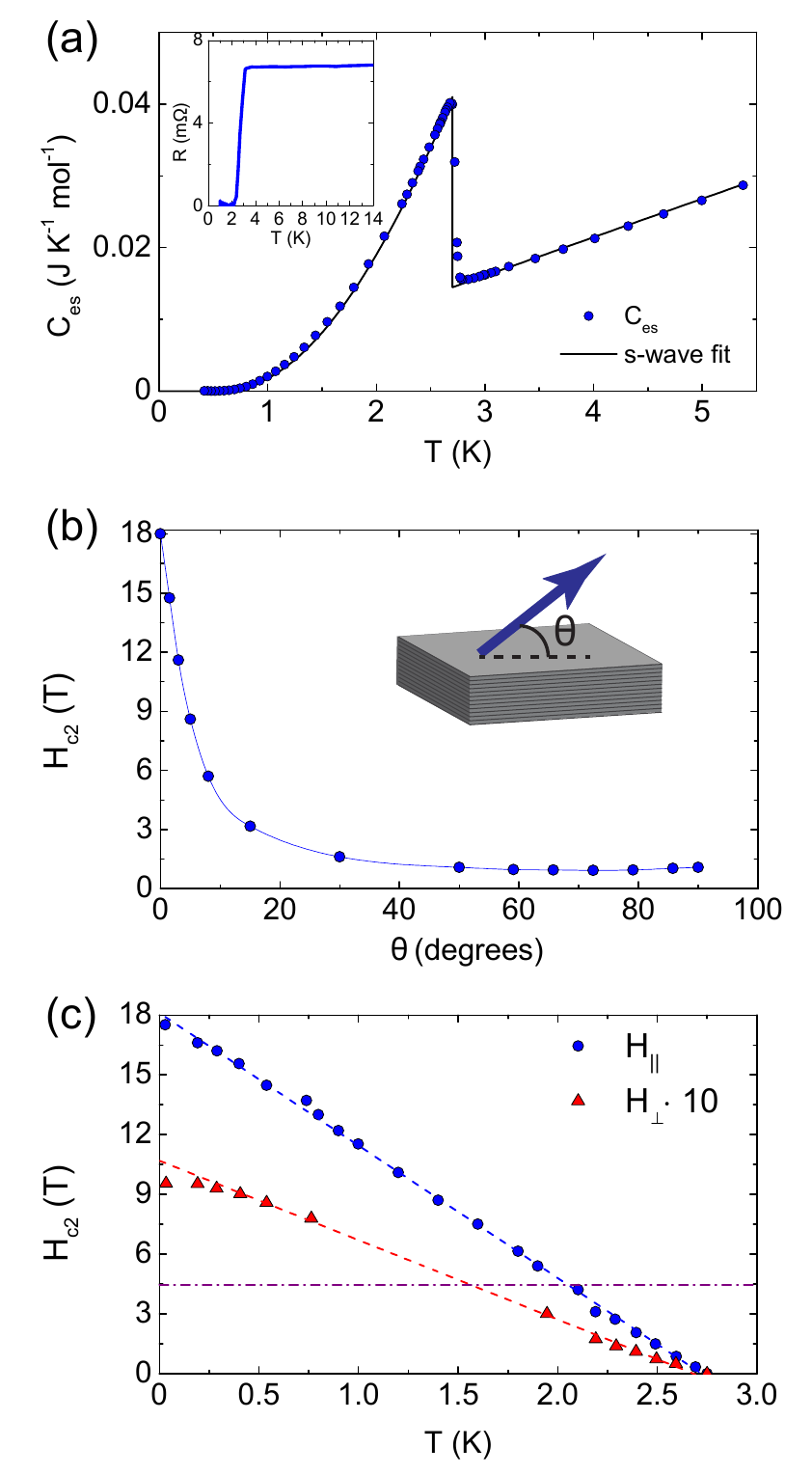}
	\caption{(a) The electronic contribution to the heat capacity C$_{es}$ is plotted as function of T. Below T$_c$ the data is in agreement with a fully-gapped s-wave model. The inset displays the resistance ($R$) as a function of temperature measured using 4-probe on the $ab$ plane. $R$ decreases monotonically until it sharply drops to zero at T$_c\simeq$2.7 K. (b) The angular dependence of H$_{C2}$ shows a large anisotropy with a factor of 18 between the in-plane and out-of-plane critical magnetic fields. (c) H$_{c2}$ as a function of temperature with H being aligned in-plane (blue) and out-of-plane (red) directions. For convenience, H$_{c2\perp}$ is multiplied by 10. While H$_{c2\perp}$ saturates at low temperatures, H$_{c2\parallel}$ remains linear down to 30 mK and up to 18 T. H$_{c2\parallel}$ exceeds the BCS Pauli limit (marked by purple dashed line) by a factor of 4.} 		
	\label{transport}
\end{figure}

Figure~\ref{transport}(a) shows $C_{es}$, the electronic part of the specific heat after the removal of the phonon contribution, as a function of the temperature {(For details, see supplementary material)}. The transition into a superconducting state below T$_c$=2.7K is clearly seen both in specific heat and resistance measurements (see inset of Fig.~\ref{transport}(a)).
The exponential behavior at low temperatures and the good agreement with an s-wave model (solid line in Fig.~\ref{transport}(a)) indicates that the system is fully-gapped.
Using the fit to a BCS model we extract a gap of $\Delta_0^{min}=0.4 \pm 0.05$ meV. {This value is in good agreement with the transverse field muon spin rotation ($\mu$SR) results shown in the supplementary material. }

To characterize the anisotropy of the superconducting state we performed magneto-transport measurements.
In Fig. \ref{transport}(b) we show the critical field, H$_{c2}$, as a function of the angle $\theta$, between the applied field and the $ab$ plane ($\theta=0$ denotes field aligned in the plane),
measured at T=30mK.
The magneto-resistance exhibits strong anisotropy, with H$_{c2}^\parallel/$H$_{c2}^\perp>17$, reflecting the quasi-2D nature of superconductivity in 4Hb-TaS$_2$. This picture is further supported by the temperature dependence of H$_{c2}^\parallel$ shown in Fig.~\ref{transport}(c). Here, we  observe an unusual linear dependance of the in-plane critical field throughout the entire range of temperature, similar to other unconventional superconductors~\cite{Paglione}.
We note that, while the parallel field exceeds the Clogston-Chandrasekhar limit, the linear dependence suggests that the Zeeman field required to align the Cooper-pair spins is much greater. On the other hand, if we interpret H$_{c2}^\parallel$ as an orbitally limited field within an anisotropic Ginzburg-Landau theory, we arrive at an out-of-plane coherence length of $\xi_c=9.8\text{\AA}$.
This length is comparable to the interlayer spacing of 1H layers. Comparing $\xi_c$ with the in-plane coherence length estimated from H$_c^\perp$, $\xi_{ab}=185.6\,\AA$, we conclude that 4Hb-TaS$_2$ is essentially a stack of weakly coupled two-dimensional superconductors.

\begin{figure}[h!]
	\centering	
	\includegraphics[width=90mm]{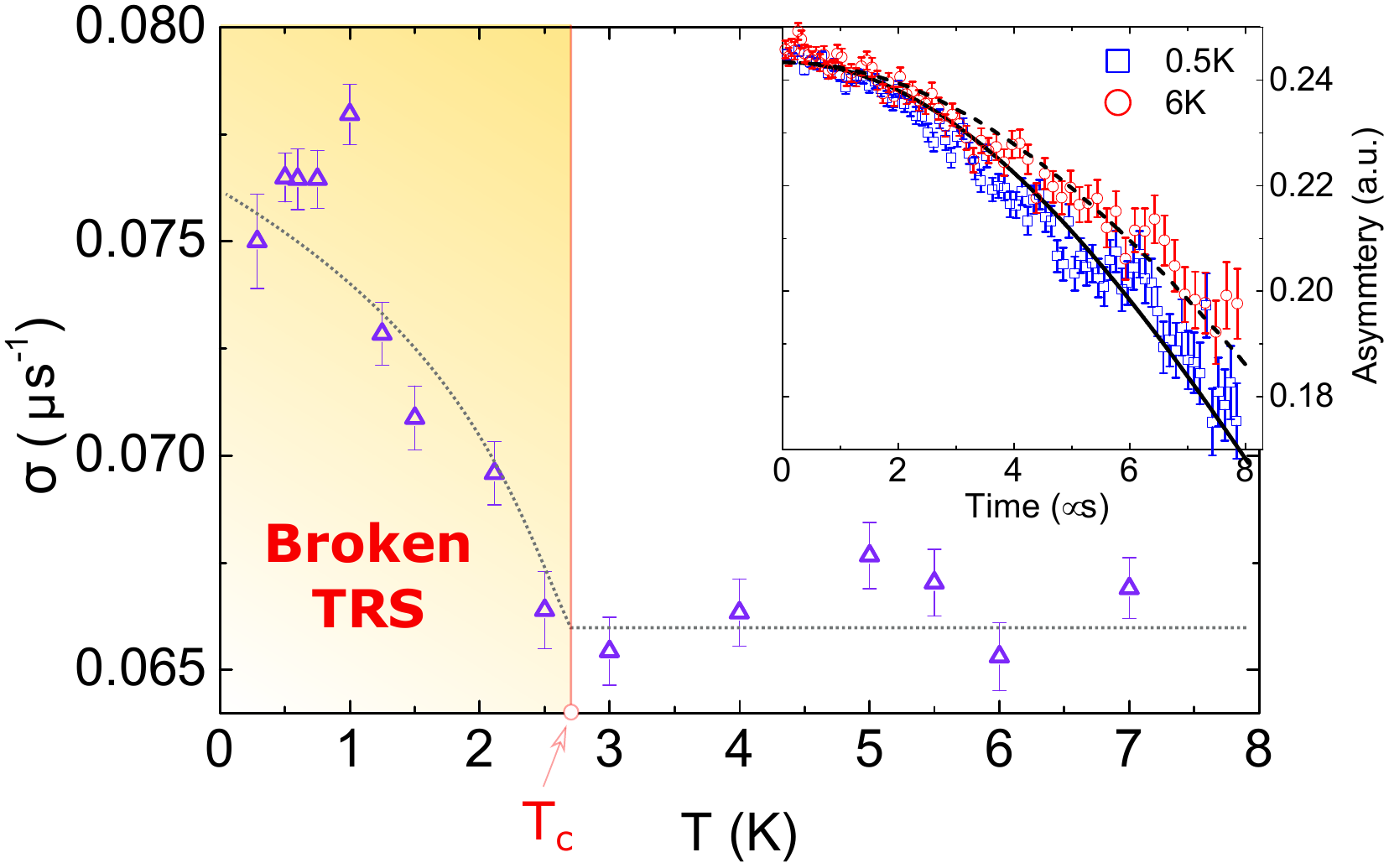}
	\caption{Zero-field muon-spin relaxation results (ZF-$\mu$SR).  A sudden increase in the muon relaxation rate, $\sigma$, is observed at T$_c$ (marked by the vertical line), marking the onset of time-reversal symmetry breaking.
The dashed line is a guide to the eye. The inset shows two ZF-$\mu$SR  spectra at two different temperatures, above and below T$_c$. The black lines are best fits to Eq. \ref{KTfunc}.}
	\label{muSR}
\end{figure}

The main result of the paper is evidence for time-reversal-symmetry breaking in the superconducting state from muon spin relaxation  ($\mu$SR) measurement. In the absence of magnetic order, the muon depolarization is a result of the randomly oriented static nuclear dipole moments and is described by the static Gaussian Kubo-Toyabe function~\cite{KT_func}
\begin{equation}
G_z(t)=\frac{1}{3}+\frac{2}{3}(1-\sigma^2t^2)\text{exp}(-\frac{1}{2}\sigma^2t^2),
\label{KTfunc}
\end{equation}
where $\sigma/\gamma_{\mu}$ is the local field distribution width and $\gamma_{\mu}$ is the muon gyromagnetic ratio. The nuclear field distribution is temperature independent.

 In Fig. \ref{muSR}, we show the temperature dependence of the muon depolarization rate in the absence of a magnetic field. We find an abrupt increase of the rate  at T = T$_c$. This is a clear indication of a spontaneous appearance of magnetic moments in the sample due to the superconducting state.
 In the inset of Fig.~\ref{muSR}, we present two representative ZF-$\mu$SR spectra, above and below T$_c$ at 6 K and 0.05 K.  Based on the increase in $\sigma$ we estimate the width of the randomly oriented magnetic field in the sample to be $\sim0.12$G.

A natural explanation for the signature of TRS breaking is chiral superconductivity.
In such a state, magnetic moments are expected to appear due to local variations in the chiral order parameter resulting, for example, from edges or defects in the sample~\cite{Ueda}.
While there are other scenarios that can account for signatures of TRS breaking within a superconducting phase, notably combinations of nearly-degenerate symmetry-distinct order parameters or frustrated interband Cooper-pair scattering, these generically happen not at $T_c$, but at a distinct lower temperature~\cite{SrPtAs_muSR}.
Thus, the most probable picture emerging from the data is that of a quasi-two-dimensional fully-gapped chiral superconductor.

Chiral superconductivity is only allowed for a multi-component gap function, in other words when the order parameter belongs to a higher-dimensional irreducible representation of the point-group symmetry. The symmetry classification of irreducible representations in the three-dimensional space group (No. 194) {was} derived in Ref. \cite{Goryo2012}. For the sake of simplicity, we discuss here a single 1H layer with point group symmetry $D_{3h}$.
Then, there is only one two-dimensional irreducible representation, referred to as $E$. Due to the three-fold rotation symmetry of a single layer, representations with equal angular momentum modulo 3 are identified. Therefore, the superconducting order parameters relevant for our situation are combinations of $p+ip$-wave and $d+id$-wave order parameters.
On a lattice, the SC order parameter in this representation can be written as
\begin{align}
&\Delta_{E}({\boldsymbol{k}})=\Delta_0[\alpha\,e_{\boldsymbol{k}}\sigma^0+(1-\alpha)o_{\boldsymbol{k}}\sigma^z]i\sigma^y\label{gap},
\end{align}
where $e_\mathbf{k}=\sum_{j=1}^{3}\omega^j\text{cos}({\boldsymbol{k}}\cdot {\boldsymbol{T}}_j)$ and $o_{\boldsymbol{k}}=\sum_{j=1}^{3}\omega^{-j}\text{sin}({\boldsymbol{k}}\cdot {\boldsymbol{T}}_j)$ are the $d$-wave and $p$-wave basis functions in the $D_{3h}$ point group. Here, we have chosen only one chirality, in other words, only Cooper pairs with positive orbital angular momentum, while their negative counterparts are given by complex conjugation.
The vectors $\pm\boldsymbol{T}_j$ point to the six nearest neighbors on the triangular Ta lattice, $\omega=\text{exp}(2\pi i/3)$, and the parameter $\alpha$ is a non-universal weight, which quantifies the mixing of the $d$-wave and $p$-wave components. Finally, $\sigma^0$ and $\sigma^i$ denote the identity and Pauli matrices. Note that due to the alternate stacking of the 1H layers, the relative phase of the order parameters in Eq.\eqref{gap} changes from layer to layer {\cite{Goryo2012}}.

\begin{figure}[h!]
	\centering	
	\includegraphics[width=0.85\linewidth]{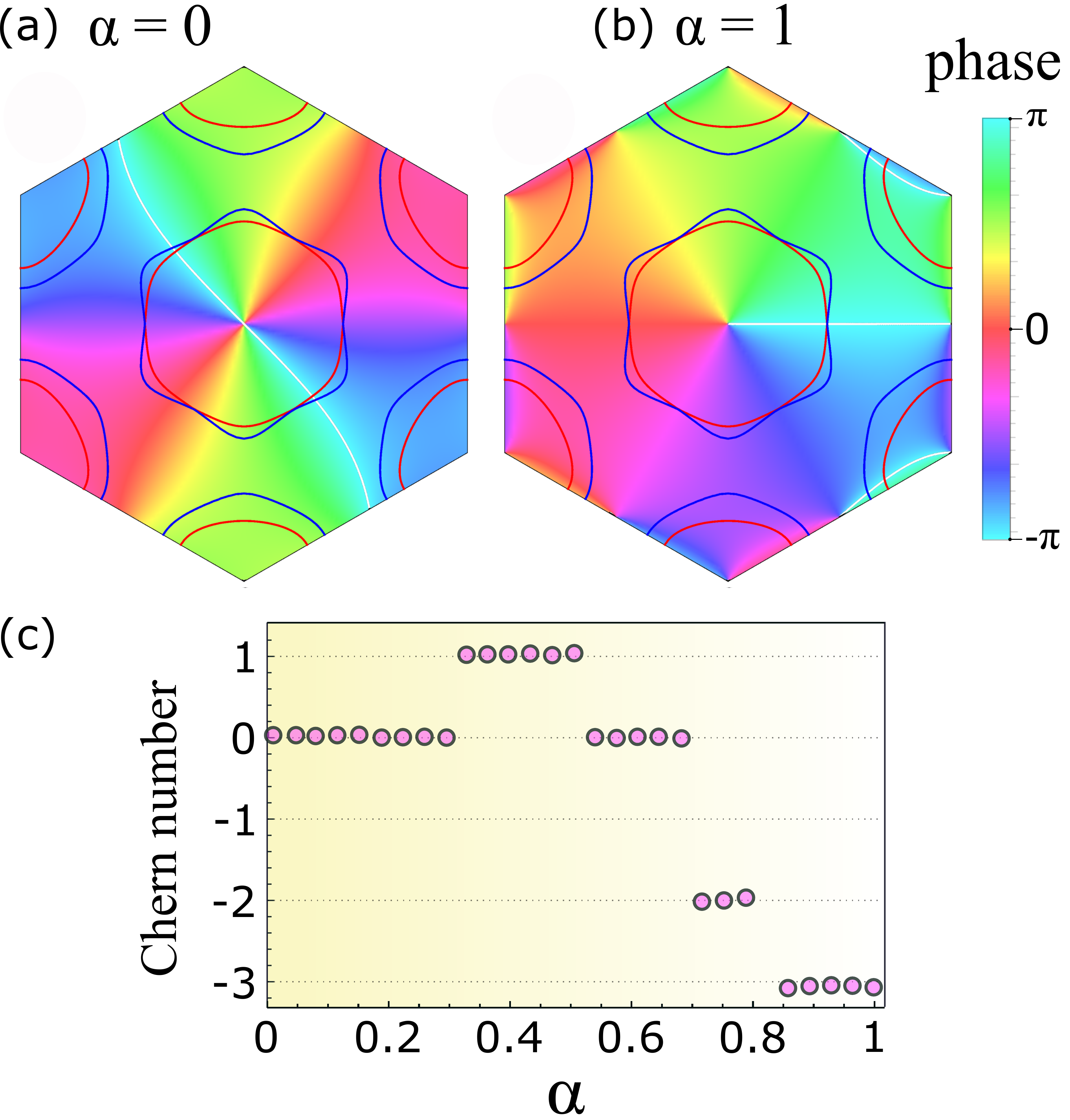}
	\caption{(a) and (b) show the phase of the gap functions $e_{\boldsymbol k}$ and $o_{\boldsymbol k}$, respectively, superimposed on the Fermi surface of 1H-TaS$_2$. (c) The corresponding Chern number as a function of the mixing parameter $\alpha$ appearing in Eq.~\eqref{gap}. Note that the values presented here were computed numerically and are not rounded to an integer.}
	\label{theory}
\end{figure}

Chiral superconductivity belongs to class D of the ten-fold classification scheme of topological states~\cite{ryu:2010}, which is trivial in three dimensions, but allows for a $\mathbb{Z}$ invariant in two dimensions. { Given the anisotropic Fermi surfaces observed in ARPES, the highly anisotropic transport properties, and the magnetic response, we consider the system as a stack of 2D chiral superconductors. That is, we compute the Chern number independently from the crystal momentum along $z$.}
Using the tight-binding model for TaS$_2$ up to the third nearest neighbor hopping~\cite{Liu2013,Law,Khodas}, we compute the 2D Chern number within a BdG Hamiltonian as a function of $\alpha$, allowing us to interpolate between the pure $d-$ and $p$-wave pairing channels.
First, we note that each one of the two non-degenerate spin bands contribute to the total Chern density.
The phase of the superconducting order parameter Eq.~\eqref{gap} is presented in Fig.~\ref{theory}.(a) and (b) for the extreme cases $\alpha = 0$ (purely $d$-wave) and $\alpha = 1$ (purely $p$-wave), respectively. Considering only one of the two relevant bands (they contribute equally), we compute the Chern number. We find $\mc C = 0$ in the limit of $\a = 0$, while for $\a = 1$ it is $\mc C = -3$. The interpolation between these two points is plotted in Fig.~\ref{theory}. (c)~\cite{Zinkl}. Note that the data points were computed numerically (not rounded to an integer) using the full BdG band structure with a mesh grid of $9.5\times10^4$ equally spaced points. Overall, we find that the Chern number of the Chiral state is highly sensitive to the mixing ratio $\alpha$. A challenging experimental goal is thus to measure a quantized thermal Hall conductance in this system.

An interesting question regards the microscopic origin of chiral superconductivity.
Phonon mediated interactions typically favor s-wave superconductivity. Unconventional pairing is expected only if strong local repulsion, which reduces the attraction in the s-wave channel, is present~\cite{Brydon}. On the other hand, an attractive interaction mediated by spin-fluctuations naturally prefers non s-wave superconductivity. From this perspective, it is interesting to understand, whether the proximity between the superconductor in the 1H layers and the Mott insulating state in the 1T layers is an essential ingredient. Our ARPES data in Fig.~\ref{Struct} (e) suggest the possibility that the Mott insulator is lightly doped due to the stacking structure, resulting in strong spin fluctuations.
Electronic pairing mediated by spin fluctuations in a quantum spin ice has been studied in Ref.\cite{She2017} for the case of a rotationally symmetric Fermi surface. {There, the authors} found that the strongest pairing channel is odd-parity with the possibility of a multi-component order parameter, consistent with the chiral superconductor found here. 
{The results presented here, thus, raise a host of theoretical questions regarding the interaction between superconductivity, charged, and neutral itinerant fermionic excitations, which invite further study.}

To summarize, we have studied 4Hb-TaS$_2$, which consists of stacked layers of 1H-TaS$_2$ that show large ASCO and 1T-TaS2$_2$, a doped Mott insulator with potentially strong spin dynamics. Thus, it naturally realizes a unique structure of stacked, two-dimensional strongly-correlated phases: a Mott insulator proposed to be a gapless spin-liquid and a two-dimensional superconductor. However, unique to this case, the two layers consist of the same chemical compound with only small difference in crystal structure. Our data suggest that the combination of these two ingredients leads to chiral superconductivity.
Thus, the relatively high T$_c\simeq2.7$K, its quasi-2D structure, the ability to grow very clean samples and the ease in which large single crystals can be grown makes this a promising platform for future study and applications. Furthermore, it is also opens new directions in the study of topological superconductivity using vdW heterostrucutres.

{\textit{Acknowledgments --} We thank Itamar Kimchi for helpful discussions.
The $\mu$SR measurements were performed at the Swiss Muon Source (S$\mu$S) at the Paul Scherrer Institute in Villigen, Switzerland.  We acknowledge the Paul Scherrer Institut, Villigen, Switzerland for provision of synchrotron radiation beam-time at beamline HRPES-SIS of the SLS. Some of this work was carried out with the support of the Diamond Light Source, instrument IO5 (proposal SI15822).
A portion of this work was performed at the National High Magnetic Field Laboratory, which is supported by the National Science Foundation Cooperative Agreement No. DMR-1644779 and the State of Florida.   This work was supported by the Israeli Science Foundation (work at the Technion under Grant No. 320/17 and work at Tel-Aviv university under Grant No. 382/17). }

\bibliography{TaS2_4Hb_SC_bib}

\begin{thebibliography}{32}%
\makeatletter
\providecommand \@ifxundefined [1]{%
 \@ifx{#1\undefined}
}%
\providecommand \@ifnum [1]{%
 \ifnum #1\expandafter \@firstoftwo
 \else \expandafter \@secondoftwo
 \fi
}%
\providecommand \@ifx [1]{%
 \ifx #1\expandafter \@firstoftwo
 \else \expandafter \@secondoftwo
 \fi
}%
\providecommand \natexlab [1]{#1}%
\providecommand \enquote  [1]{``#1''}%
\providecommand \bibnamefont  [1]{#1}%
\providecommand \bibfnamefont [1]{#1}%
\providecommand \citenamefont [1]{#1}%
\providecommand \href@noop [0]{\@secondoftwo}%
\providecommand \href [0]{\begingroup \@sanitize@url \@href}%
\providecommand \@href[1]{\@@startlink{#1}\@@href}%
\providecommand \@@href[1]{\endgroup#1\@@endlink}%
\providecommand \@sanitize@url [0]{\catcode `\\12\catcode `\$12\catcode
  `\&12\catcode `\#12\catcode `\^12\catcode `\_12\catcode `\%12\relax}%
\providecommand \@@startlink[1]{}%
\providecommand \@@endlink[0]{}%
\providecommand \url  [0]{\begingroup\@sanitize@url \@url }%
\providecommand \@url [1]{\endgroup\@href {#1}{\urlprefix }}%
\providecommand \urlprefix  [0]{URL }%
\providecommand \Eprint [0]{\href }%
\providecommand \doibase [0]{http://dx.doi.org/}%
\providecommand \selectlanguage [0]{\@gobble}%
\providecommand \bibinfo  [0]{\@secondoftwo}%
\providecommand \bibfield  [0]{\@secondoftwo}%
\providecommand \translation [1]{[#1]}%
\providecommand \BibitemOpen [0]{}%
\providecommand \bibitemStop [0]{}%
\providecommand \bibitemNoStop [0]{.\EOS\space}%
\providecommand \EOS [0]{\spacefactor3000\relax}%
\providecommand \BibitemShut  [1]{\csname bibitem#1\endcsname}%
\let\auto@bib@innerbib\@empty
\bibitem [{\citenamefont {Geim}\ and\ \citenamefont
  {Grigorieva}(2013)}]{Geim_2D}%
  \BibitemOpen
  \bibfield  {author} {\bibinfo {author} {\bibfnamefont {A.~K.}\ \bibnamefont
  {Geim}}\ and\ \bibinfo {author} {\bibfnamefont {I.~V.}\ \bibnamefont
  {Grigorieva}},\ }\href@noop {} {\bibfield  {journal} {\bibinfo  {journal}
  {Nature}\ }\textbf {\bibinfo {volume} {499}},\ \bibinfo {pages} {419}
  (\bibinfo {year} {2013})}\BibitemShut {NoStop}%
\bibitem [{\citenamefont {Novoselov}\ \emph {et~al.}(2016)\citenamefont
  {Novoselov}, \citenamefont {Mishchenko}, \citenamefont {Carvalho},\ and\
  \citenamefont {Castro~Neto}}]{novoselov:2016}%
  \BibitemOpen
  \bibfield  {author} {\bibinfo {author} {\bibfnamefont {K.~S.}\ \bibnamefont
  {Novoselov}}, \bibinfo {author} {\bibfnamefont {A.}~\bibnamefont
  {Mishchenko}}, \bibinfo {author} {\bibfnamefont {A.}~\bibnamefont
  {Carvalho}}, \ and\ \bibinfo {author} {\bibfnamefont {A.~H.}\ \bibnamefont
  {Castro~Neto}},\ }\href@noop {} {\bibfield  {journal} {\bibinfo  {journal}
  {Science}\ }\textbf {\bibinfo {volume} {353}},\ \bibinfo {pages} {461}
  (\bibinfo {year} {2016})}\BibitemShut {NoStop}%
\bibitem [{\citenamefont {Cao}\ \emph {et~al.}(2018{\natexlab{a}})\citenamefont
  {Cao}, \citenamefont {Fatemi}, \citenamefont {Demir}, \citenamefont {Fang},
  \citenamefont {Tomarken}, \citenamefont {Luo}, \citenamefont
  {Sanchez-Yamagishi}, \citenamefont {Watanabe}, \citenamefont {Taniguchi},
  \citenamefont {Kaxiras} \emph {et~al.}}]{Pablo1}%
  \BibitemOpen
  \bibfield  {author} {\bibinfo {author} {\bibfnamefont {Y.}~\bibnamefont
  {Cao}}, \bibinfo {author} {\bibfnamefont {V.}~\bibnamefont {Fatemi}},
  \bibinfo {author} {\bibfnamefont {A.}~\bibnamefont {Demir}}, \bibinfo
  {author} {\bibfnamefont {S.}~\bibnamefont {Fang}}, \bibinfo {author}
  {\bibfnamefont {S.~L.}\ \bibnamefont {Tomarken}}, \bibinfo {author}
  {\bibfnamefont {J.~Y.}\ \bibnamefont {Luo}}, \bibinfo {author} {\bibfnamefont
  {J.~D.}\ \bibnamefont {Sanchez-Yamagishi}}, \bibinfo {author} {\bibfnamefont
  {K.}~\bibnamefont {Watanabe}}, \bibinfo {author} {\bibfnamefont
  {T.}~\bibnamefont {Taniguchi}}, \bibinfo {author} {\bibfnamefont
  {E.}~\bibnamefont {Kaxiras}},  \emph {et~al.},\ }\href@noop {} {\bibfield
  {journal} {\bibinfo  {journal} {Nature}\ }\textbf {\bibinfo {volume} {556}},\
  \bibinfo {pages} {80} (\bibinfo {year} {2018}{\natexlab{a}})}\BibitemShut
  {NoStop}%
\bibitem [{\citenamefont {Cao}\ \emph {et~al.}(2018{\natexlab{b}})\citenamefont
  {Cao}, \citenamefont {Fatemi}, \citenamefont {Fang}, \citenamefont
  {Watanabe}, \citenamefont {Taniguchi}, \citenamefont {Kaxiras},\ and\
  \citenamefont {Jarillo-Herrero}}]{Pablo2}%
  \BibitemOpen
  \bibfield  {author} {\bibinfo {author} {\bibfnamefont {Y.}~\bibnamefont
  {Cao}}, \bibinfo {author} {\bibfnamefont {V.}~\bibnamefont {Fatemi}},
  \bibinfo {author} {\bibfnamefont {S.}~\bibnamefont {Fang}}, \bibinfo {author}
  {\bibfnamefont {K.}~\bibnamefont {Watanabe}}, \bibinfo {author}
  {\bibfnamefont {T.}~\bibnamefont {Taniguchi}}, \bibinfo {author}
  {\bibfnamefont {E.}~\bibnamefont {Kaxiras}}, \ and\ \bibinfo {author}
  {\bibfnamefont {P.}~\bibnamefont {Jarillo-Herrero}},\ }\href@noop {}
  {\bibfield  {journal} {\bibinfo  {journal} {Nature}\ }\textbf {\bibinfo
  {volume} {556}},\ \bibinfo {pages} {43} (\bibinfo {year}
  {2018}{\natexlab{b}})}\BibitemShut {NoStop}%
\bibitem [{\citenamefont {Law}\ and\ \citenamefont {Lee}(2017)}]{Lee}%
  \BibitemOpen
  \bibfield  {author} {\bibinfo {author} {\bibfnamefont {K.~T.}\ \bibnamefont
  {Law}}\ and\ \bibinfo {author} {\bibfnamefont {P.~A.}\ \bibnamefont {Lee}},\
  }\href {http://www.pnas.org/content/early/2017/06/19/1706769114.abstract N2 -
  1T-TaS2 is unique among transition metal dichalcogenides in that it is
  understood to be a correlation-driven insulator, where the unpaired electron
  in a 13-site cluster experiences enough correlation to form a Mott insulator.
  We argue, based on existing data, that this well-known material should be
  considered as a quantum spin liquid, either a fully gapped Z2 spin liquid or
  a Dirac spin liquid. We discuss the exotic states that emerge upon doping and
  propose further experimental probes.} {\bibfield  {journal} {\bibinfo
  {journal} {Proceedings of the National Academy of Sciences}\ }\textbf
  {\bibinfo {volume} {114}},\ \bibinfo {pages} {6996} (\bibinfo {year}
  {2017})}\BibitemShut {NoStop}%
\bibitem [{\citenamefont {Ribak}\ \emph {et~al.}(2017)\citenamefont {Ribak},
  \citenamefont {Silber}, \citenamefont {Baines}, \citenamefont {Chashka},
  \citenamefont {Salman}, \citenamefont {Dagan},\ and\ \citenamefont
  {Kanigel}}]{Ribak2017}%
  \BibitemOpen
  \bibfield  {author} {\bibinfo {author} {\bibfnamefont {A.}~\bibnamefont
  {Ribak}}, \bibinfo {author} {\bibfnamefont {I.}~\bibnamefont {Silber}},
  \bibinfo {author} {\bibfnamefont {C.}~\bibnamefont {Baines}}, \bibinfo
  {author} {\bibfnamefont {K.}~\bibnamefont {Chashka}}, \bibinfo {author}
  {\bibfnamefont {Z.}~\bibnamefont {Salman}}, \bibinfo {author} {\bibfnamefont
  {Y.}~\bibnamefont {Dagan}}, \ and\ \bibinfo {author} {\bibfnamefont
  {A.}~\bibnamefont {Kanigel}},\ }\href
  {https://link.aps.org/doi/10.1103/PhysRevB.96.195131} {\bibfield  {journal}
  {\bibinfo  {journal} {Phys. Rev. B}\ }\textbf {\bibinfo {volume} {96}},\
  \bibinfo {pages} {195131} (\bibinfo {year} {2017})}\BibitemShut {NoStop}%
\bibitem [{\citenamefont {Yang}\ \emph {et~al.}(2018)\citenamefont {Yang},
  \citenamefont {Fang}, \citenamefont {Fatemi}, \citenamefont {Ruhman},
  \citenamefont {Navarro-Moratalla}, \citenamefont {Watanabe}, \citenamefont
  {Taniguchi}, \citenamefont {Kaxiras},\ and\ \citenamefont
  {Jarillo-Herrero}}]{yang2018enhanced}%
  \BibitemOpen
  \bibfield  {author} {\bibinfo {author} {\bibfnamefont {Y.}~\bibnamefont
  {Yang}}, \bibinfo {author} {\bibfnamefont {S.}~\bibnamefont {Fang}}, \bibinfo
  {author} {\bibfnamefont {V.}~\bibnamefont {Fatemi}}, \bibinfo {author}
  {\bibfnamefont {J.}~\bibnamefont {Ruhman}}, \bibinfo {author} {\bibfnamefont
  {E.}~\bibnamefont {Navarro-Moratalla}}, \bibinfo {author} {\bibfnamefont
  {K.}~\bibnamefont {Watanabe}}, \bibinfo {author} {\bibfnamefont
  {T.}~\bibnamefont {Taniguchi}}, \bibinfo {author} {\bibfnamefont
  {E.}~\bibnamefont {Kaxiras}}, \ and\ \bibinfo {author} {\bibfnamefont
  {P.}~\bibnamefont {Jarillo-Herrero}},\ }\href@noop {} {\bibfield  {journal}
  {\bibinfo  {journal} {Physical Review B}\ }\textbf {\bibinfo {volume} {98}},\
  \bibinfo {pages} {035203} (\bibinfo {year} {2018})}\BibitemShut {NoStop}%
\bibitem [{\citenamefont {Read}\ and\ \citenamefont {Green}(2000)}]{ReadGreen}%
  \BibitemOpen
  \bibfield  {author} {\bibinfo {author} {\bibfnamefont {N.}~\bibnamefont
  {Read}}\ and\ \bibinfo {author} {\bibfnamefont {D.}~\bibnamefont {Green}},\
  }\href {\doibase 10.1103/PhysRevB.61.10267} {\bibfield  {journal} {\bibinfo
  {journal} {Phys. Rev. B}\ }\textbf {\bibinfo {volume} {61}},\ \bibinfo
  {pages} {10267} (\bibinfo {year} {2000})}\BibitemShut {NoStop}%
\bibitem [{\citenamefont {Sau}\ \emph {et~al.}(2010)\citenamefont {Sau},
  \citenamefont {Lutchyn}, \citenamefont {Tewari},\ and\ \citenamefont
  {Sarma}}]{sau2010generic}%
  \BibitemOpen
  \bibfield  {author} {\bibinfo {author} {\bibfnamefont {J.~D.}\ \bibnamefont
  {Sau}}, \bibinfo {author} {\bibfnamefont {R.~M.}\ \bibnamefont {Lutchyn}},
  \bibinfo {author} {\bibfnamefont {S.}~\bibnamefont {Tewari}}, \ and\ \bibinfo
  {author} {\bibfnamefont {S.~D.}\ \bibnamefont {Sarma}},\ }\href@noop {}
  {\bibfield  {journal} {\bibinfo  {journal} {Physical review letters}\
  }\textbf {\bibinfo {volume} {104}},\ \bibinfo {pages} {040502} (\bibinfo
  {year} {2010})}\BibitemShut {NoStop}%
\bibitem [{\citenamefont {Nayak}\ \emph {et~al.}(2008)\citenamefont {Nayak},
  \citenamefont {Simon}, \citenamefont {Stern}, \citenamefont {Freedman},\ and\
  \citenamefont {Das~Sarma}}]{NayakReview}%
  \BibitemOpen
  \bibfield  {author} {\bibinfo {author} {\bibfnamefont {C.}~\bibnamefont
  {Nayak}}, \bibinfo {author} {\bibfnamefont {S.~H.}\ \bibnamefont {Simon}},
  \bibinfo {author} {\bibfnamefont {A.}~\bibnamefont {Stern}}, \bibinfo
  {author} {\bibfnamefont {M.}~\bibnamefont {Freedman}}, \ and\ \bibinfo
  {author} {\bibfnamefont {S.}~\bibnamefont {Das~Sarma}},\ }\href
  {https://link.aps.org/doi/10.1103/RevModPhys.80.1083} {\bibfield  {journal}
  {\bibinfo  {journal} {Rev. Mod. Phys.}\ }\textbf {\bibinfo {volume} {80}},\
  \bibinfo {pages} {1083} (\bibinfo {year} {2008})}\BibitemShut {NoStop}%
\bibitem [{\citenamefont {Sigrist}\ and\ \citenamefont {Ueda}(1991)}]{Ueda}%
  \BibitemOpen
  \bibfield  {author} {\bibinfo {author} {\bibfnamefont {M.}~\bibnamefont
  {Sigrist}}\ and\ \bibinfo {author} {\bibfnamefont {K.}~\bibnamefont {Ueda}},\
  }\href {\doibase 10.1103/RevModPhys.63.239} {\bibfield  {journal} {\bibinfo
  {journal} {Rev. Mod. Phys.}\ }\textbf {\bibinfo {volume} {63}},\ \bibinfo
  {pages} {239} (\bibinfo {year} {1991})}\BibitemShut {NoStop}%
\bibitem [{\citenamefont {Luke}\ \emph {et~al.}(1998)\citenamefont {Luke},
  \citenamefont {Fudamoto}, \citenamefont {Kojima}, \citenamefont {Larkin},
  \citenamefont {Merrin}, \citenamefont {Nachumi}, \citenamefont {Uemura},
  \citenamefont {Maeno}, \citenamefont {Mao}, \citenamefont {Mori},
  \citenamefont {Nakamura},\ and\ \citenamefont {Sigrist}}]{Luke_SRO}%
  \BibitemOpen
  \bibfield  {author} {\bibinfo {author} {\bibfnamefont {G.~M.}\ \bibnamefont
  {Luke}}, \bibinfo {author} {\bibfnamefont {Y.}~\bibnamefont {Fudamoto}},
  \bibinfo {author} {\bibfnamefont {K.~M.}\ \bibnamefont {Kojima}}, \bibinfo
  {author} {\bibfnamefont {M.~I.}\ \bibnamefont {Larkin}}, \bibinfo {author}
  {\bibfnamefont {J.}~\bibnamefont {Merrin}}, \bibinfo {author} {\bibfnamefont
  {B.}~\bibnamefont {Nachumi}}, \bibinfo {author} {\bibfnamefont {Y.~J.}\
  \bibnamefont {Uemura}}, \bibinfo {author} {\bibfnamefont {Y.}~\bibnamefont
  {Maeno}}, \bibinfo {author} {\bibfnamefont {Z.~Q.}\ \bibnamefont {Mao}},
  \bibinfo {author} {\bibfnamefont {Y.}~\bibnamefont {Mori}}, \bibinfo {author}
  {\bibfnamefont {H.}~\bibnamefont {Nakamura}}, \ and\ \bibinfo {author}
  {\bibfnamefont {M.}~\bibnamefont {Sigrist}},\ }\href
  {https://doi.org/10.1038/29038} {\bibfield  {journal} {\bibinfo  {journal}
  {Nature}\ }\textbf {\bibinfo {volume} {394}},\ \bibinfo {pages} {558}
  (\bibinfo {year} {1998})}\BibitemShut {NoStop}%
\bibitem [{\citenamefont {Schemm}\ \emph {et~al.}(2014)\citenamefont {Schemm},
  \citenamefont {Gannon}, \citenamefont {Wishne}, \citenamefont {Halperin},\
  and\ \citenamefont {Kapitulnik}}]{schemm2014observation}%
  \BibitemOpen
  \bibfield  {author} {\bibinfo {author} {\bibfnamefont {E.}~\bibnamefont
  {Schemm}}, \bibinfo {author} {\bibfnamefont {W.}~\bibnamefont {Gannon}},
  \bibinfo {author} {\bibfnamefont {C.}~\bibnamefont {Wishne}}, \bibinfo
  {author} {\bibfnamefont {W.~P.}\ \bibnamefont {Halperin}}, \ and\ \bibinfo
  {author} {\bibfnamefont {A.}~\bibnamefont {Kapitulnik}},\ }\href@noop {}
  {\bibfield  {journal} {\bibinfo  {journal} {Science}\ }\textbf {\bibinfo
  {volume} {345}},\ \bibinfo {pages} {190} (\bibinfo {year}
  {2014})}\BibitemShut {NoStop}%
\bibitem [{\citenamefont {Kawasaki}\ \emph {et~al.}(2014)\citenamefont
  {Kawasaki}, \citenamefont {Watanabe}, \citenamefont {Hillier},\ and\
  \citenamefont {Aoki}}]{kawasaki2014}%
  \BibitemOpen
  \bibfield  {author} {\bibinfo {author} {\bibfnamefont {I.}~\bibnamefont
  {Kawasaki}}, \bibinfo {author} {\bibfnamefont {I.}~\bibnamefont {Watanabe}},
  \bibinfo {author} {\bibfnamefont {A.}~\bibnamefont {Hillier}}, \ and\
  \bibinfo {author} {\bibfnamefont {D.}~\bibnamefont {Aoki}},\ }\href {\doibase
  10.7566/jpsj.83.094720} {\bibfield  {journal} {\bibinfo  {journal} {J. Phys.
  Soc. Jpn.}\ }\textbf {\bibinfo {volume} {83}},\ \bibinfo {pages} {094720}
  (\bibinfo {year} {2014})}\BibitemShut {NoStop}%
\bibitem [{\citenamefont {Biswas}\ \emph {et~al.}(2013)\citenamefont {Biswas},
  \citenamefont {Luetkens}, \citenamefont {Neupert}, \citenamefont {Stürzer},
  \citenamefont {Baines}, \citenamefont {Pascua}, \citenamefont {Schnyder},
  \citenamefont {Fischer}, \citenamefont {Goryo}, \citenamefont {Lees},
  \citenamefont {Maeter}, \citenamefont {Brückner}, \citenamefont {Klauss},
  \citenamefont {Nicklas}, \citenamefont {Baker}, \citenamefont {Hillier},
  \citenamefont {Sigrist}, \citenamefont {Amato},\ and\ \citenamefont
  {Johrendt}}]{SrPtAs_muSR}%
  \BibitemOpen
  \bibfield  {author} {\bibinfo {author} {\bibfnamefont {P.~K.}\ \bibnamefont
  {Biswas}}, \bibinfo {author} {\bibfnamefont {H.}~\bibnamefont {Luetkens}},
  \bibinfo {author} {\bibfnamefont {T.}~\bibnamefont {Neupert}}, \bibinfo
  {author} {\bibfnamefont {T.}~\bibnamefont {Stürzer}}, \bibinfo {author}
  {\bibfnamefont {C.}~\bibnamefont {Baines}}, \bibinfo {author} {\bibfnamefont
  {G.}~\bibnamefont {Pascua}}, \bibinfo {author} {\bibfnamefont {A.~P.}\
  \bibnamefont {Schnyder}}, \bibinfo {author} {\bibfnamefont {M.~H.}\
  \bibnamefont {Fischer}}, \bibinfo {author} {\bibfnamefont {J.}~\bibnamefont
  {Goryo}}, \bibinfo {author} {\bibfnamefont {M.~R.}\ \bibnamefont {Lees}},
  \bibinfo {author} {\bibfnamefont {H.}~\bibnamefont {Maeter}}, \bibinfo
  {author} {\bibfnamefont {F.}~\bibnamefont {Brückner}}, \bibinfo {author}
  {\bibfnamefont {H.-H.}\ \bibnamefont {Klauss}}, \bibinfo {author}
  {\bibfnamefont {M.}~\bibnamefont {Nicklas}}, \bibinfo {author} {\bibfnamefont
  {P.~J.}\ \bibnamefont {Baker}}, \bibinfo {author} {\bibfnamefont {A.~D.}\
  \bibnamefont {Hillier}}, \bibinfo {author} {\bibfnamefont {M.}~\bibnamefont
  {Sigrist}}, \bibinfo {author} {\bibfnamefont {A.}~\bibnamefont {Amato}}, \
  and\ \bibinfo {author} {\bibfnamefont {D.}~\bibnamefont {Johrendt}},\ }\href
  {https://link.aps.org/doi/10.1103/PhysRevB.87.180503} {\bibfield  {journal}
  {\bibinfo  {journal} {Phys. Rev. B}\ }\textbf {\bibinfo {volume} {87}},\
  \bibinfo {pages} {180503} (\bibinfo {year} {2013})}\BibitemShut {NoStop}%
\bibitem [{\citenamefont {Fischer}\ \emph {et~al.}(2014)\citenamefont
  {Fischer}, \citenamefont {Neupert}, \citenamefont {Platt}, \citenamefont
  {Schnyder}, \citenamefont {Hanke}, \citenamefont {Goryo}, \citenamefont
  {Thomale},\ and\ \citenamefont {Sigrist}}]{Fischer2014}%
  \BibitemOpen
  \bibfield  {author} {\bibinfo {author} {\bibfnamefont {M.~H.}\ \bibnamefont
  {Fischer}}, \bibinfo {author} {\bibfnamefont {T.}~\bibnamefont {Neupert}},
  \bibinfo {author} {\bibfnamefont {C.}~\bibnamefont {Platt}}, \bibinfo
  {author} {\bibfnamefont {A.~P.}\ \bibnamefont {Schnyder}}, \bibinfo {author}
  {\bibfnamefont {W.}~\bibnamefont {Hanke}}, \bibinfo {author} {\bibfnamefont
  {J.}~\bibnamefont {Goryo}}, \bibinfo {author} {\bibfnamefont
  {R.}~\bibnamefont {Thomale}}, \ and\ \bibinfo {author} {\bibfnamefont
  {M.}~\bibnamefont {Sigrist}},\ }\href {\doibase 10.1103/PhysRevB.89.020509}
  {\bibfield  {journal} {\bibinfo  {journal} {Phys. Rev. B}\ }\textbf {\bibinfo
  {volume} {89}},\ \bibinfo {pages} {020509} (\bibinfo {year}
  {2014})}\BibitemShut {NoStop}%
\bibitem [{\citenamefont {Hassinger}\ \emph {et~al.}(2017)\citenamefont
  {Hassinger}, \citenamefont {Bourgeois-Hope}, \citenamefont {Taniguchi},
  \citenamefont {René~de Cotret}, \citenamefont {Grissonnanche}, \citenamefont
  {Anwar}, \citenamefont {Maeno}, \citenamefont {Doiron-Leyraud},\ and\
  \citenamefont {Taillefer}}]{Hassinger2017}%
  \BibitemOpen
  \bibfield  {author} {\bibinfo {author} {\bibfnamefont {E.}~\bibnamefont
  {Hassinger}}, \bibinfo {author} {\bibfnamefont {P.}~\bibnamefont
  {Bourgeois-Hope}}, \bibinfo {author} {\bibfnamefont {H.}~\bibnamefont
  {Taniguchi}}, \bibinfo {author} {\bibfnamefont {S.}~\bibnamefont {René~de
  Cotret}}, \bibinfo {author} {\bibfnamefont {G.}~\bibnamefont
  {Grissonnanche}}, \bibinfo {author} {\bibfnamefont {M.~S.}\ \bibnamefont
  {Anwar}}, \bibinfo {author} {\bibfnamefont {Y.}~\bibnamefont {Maeno}},
  \bibinfo {author} {\bibfnamefont {N.}~\bibnamefont {Doiron-Leyraud}}, \ and\
  \bibinfo {author} {\bibfnamefont {L.}~\bibnamefont {Taillefer}},\ }\href
  {https://link.aps.org/doi/10.1103/PhysRevX.7.011032} {\bibfield  {journal}
  {\bibinfo  {journal} {Phys. Rev. X}\ }\textbf {\bibinfo {volume} {7}},\
  \bibinfo {pages} {011032} (\bibinfo {year} {2017})}\BibitemShut {NoStop}%
\bibitem [{\citenamefont {Schemm}\ \emph {et~al.}(2017)\citenamefont {Schemm},
  \citenamefont {Levenson-Falk},\ and\ \citenamefont
  {Kapitulnik}}]{Schemm2017}%
  \BibitemOpen
  \bibfield  {author} {\bibinfo {author} {\bibfnamefont {E.~R.}\ \bibnamefont
  {Schemm}}, \bibinfo {author} {\bibfnamefont {E.~M.}\ \bibnamefont
  {Levenson-Falk}}, \ and\ \bibinfo {author} {\bibfnamefont {A.}~\bibnamefont
  {Kapitulnik}},\ }\href
  {http://www.sciencedirect.com/science/article/pii/S092145341630257X}
  {\bibfield  {journal} {\bibinfo  {journal} {Physica C: Superconductivity and
  its Applications}\ }\textbf {\bibinfo {volume} {535}},\ \bibinfo {pages} {13}
  (\bibinfo {year} {2017})}\BibitemShut {NoStop}%
\bibitem [{\citenamefont {Fazekas}\ and\ \citenamefont
  {Tosatti}(1980)}]{Fazekas1980}%
  \BibitemOpen
  \bibfield  {author} {\bibinfo {author} {\bibfnamefont {P.}~\bibnamefont
  {Fazekas}}\ and\ \bibinfo {author} {\bibfnamefont {E.}~\bibnamefont
  {Tosatti}},\ }\href
  {http://www.sciencedirect.com/science/article/pii/0378436380902296}
  {\bibfield  {journal} {\bibinfo  {journal} {Physica B+C}\ }\textbf {\bibinfo
  {volume} {99}},\ \bibinfo {pages} {183} (\bibinfo {year} {1980})}\BibitemShut
  {NoStop}%
\bibitem [{\citenamefont {Murayama}\ \emph {et~al.}(2018)\citenamefont
  {Murayama}, \citenamefont {Sato}, \citenamefont {Xing}, \citenamefont
  {Taniguchi}, \citenamefont {Kasahara}, \citenamefont {Kasahara},
  \citenamefont {Yoshida}, \citenamefont {Iwasa},\ and\ \citenamefont
  {Matsuda}}]{Matsuda_thermal}%
  \BibitemOpen
  \bibfield  {author} {\bibinfo {author} {\bibfnamefont {H.}~\bibnamefont
  {Murayama}}, \bibinfo {author} {\bibfnamefont {Y.}~\bibnamefont {Sato}},
  \bibinfo {author} {\bibfnamefont {X.~Z.}\ \bibnamefont {Xing}}, \bibinfo
  {author} {\bibfnamefont {T.}~\bibnamefont {Taniguchi}}, \bibinfo {author}
  {\bibfnamefont {S.}~\bibnamefont {Kasahara}}, \bibinfo {author}
  {\bibfnamefont {Y.}~\bibnamefont {Kasahara}}, \bibinfo {author}
  {\bibfnamefont {M.}~\bibnamefont {Yoshida}}, \bibinfo {author} {\bibfnamefont
  {Y.}~\bibnamefont {Iwasa}}, \ and\ \bibinfo {author} {\bibfnamefont
  {Y.}~\bibnamefont {Matsuda}},\ }\href@noop {} {\bibfield  {journal} {\bibinfo
   {journal} {arXiv}\ }\textbf {\bibinfo {volume} {1803.06100}} (\bibinfo
  {year} {2018})}\BibitemShut {NoStop}%
\bibitem [{\citenamefont {Di~Salvo}\ \emph {et~al.}(1973)\citenamefont
  {Di~Salvo}, \citenamefont {Bagley}, \citenamefont {Voorhoeve},\ and\
  \citenamefont {Waszczak}}]{DiSalvo_4Hb}%
  \BibitemOpen
  \bibfield  {author} {\bibinfo {author} {\bibfnamefont {F.~J.}\ \bibnamefont
  {Di~Salvo}}, \bibinfo {author} {\bibfnamefont {B.~G.}\ \bibnamefont
  {Bagley}}, \bibinfo {author} {\bibfnamefont {J.~M.}\ \bibnamefont
  {Voorhoeve}}, \ and\ \bibinfo {author} {\bibfnamefont {J.~V.}\ \bibnamefont
  {Waszczak}},\ }\href
  {http://www.sciencedirect.com/science/article/pii/S0022369773800344}
  {\bibfield  {journal} {\bibinfo  {journal} {Journal of Physics and Chemistry
  of Solids}\ }\textbf {\bibinfo {volume} {34}},\ \bibinfo {pages} {1357}
  (\bibinfo {year} {1973})}\BibitemShut {NoStop}%
\bibitem [{\citenamefont {Hughes}\ and\ \citenamefont
  {Scarfe}(1995)}]{TaS2-1T_4H-XPS}%
  \BibitemOpen
  \bibfield  {author} {\bibinfo {author} {\bibfnamefont {H.~P.}\ \bibnamefont
  {Hughes}}\ and\ \bibinfo {author} {\bibfnamefont {J.~A.}\ \bibnamefont
  {Scarfe}},\ }\href {https://link.aps.org/doi/10.1103/PhysRevLett.74.3069}
  {\bibfield  {journal} {\bibinfo  {journal} {Phys. Rev. Lett.}\ }\textbf
  {\bibinfo {volume} {74}},\ \bibinfo {pages} {3069} (\bibinfo {year}
  {1995})}\BibitemShut {NoStop}%
\bibitem [{\citenamefont {Butch}\ \emph {et~al.}(2011)\citenamefont {Butch},
  \citenamefont {Syers}, \citenamefont {Kirshenbaum}, \citenamefont {Hope},\
  and\ \citenamefont {Paglione}}]{Paglione}%
  \BibitemOpen
  \bibfield  {author} {\bibinfo {author} {\bibfnamefont {N.~P.}\ \bibnamefont
  {Butch}}, \bibinfo {author} {\bibfnamefont {P.}~\bibnamefont {Syers}},
  \bibinfo {author} {\bibfnamefont {K.}~\bibnamefont {Kirshenbaum}}, \bibinfo
  {author} {\bibfnamefont {A.~P.}\ \bibnamefont {Hope}}, \ and\ \bibinfo
  {author} {\bibfnamefont {J.}~\bibnamefont {Paglione}},\ }\href {\doibase
  10.1103/PhysRevB.84.220504} {\bibfield  {journal} {\bibinfo  {journal} {Phys.
  Rev. B}\ }\textbf {\bibinfo {volume} {84}},\ \bibinfo {pages} {220504}
  (\bibinfo {year} {2011})}\BibitemShut {NoStop}%
\bibitem [{\citenamefont {Hayano}\ \emph {et~al.}(1979)\citenamefont {Hayano},
  \citenamefont {Uemura}, \citenamefont {Imazato}, \citenamefont {Nishida},
  \citenamefont {Yamazaki},\ and\ \citenamefont {Kubo}}]{KT_func}%
  \BibitemOpen
  \bibfield  {author} {\bibinfo {author} {\bibfnamefont {R.~S.}\ \bibnamefont
  {Hayano}}, \bibinfo {author} {\bibfnamefont {Y.~J.}\ \bibnamefont {Uemura}},
  \bibinfo {author} {\bibfnamefont {J.}~\bibnamefont {Imazato}}, \bibinfo
  {author} {\bibfnamefont {N.}~\bibnamefont {Nishida}}, \bibinfo {author}
  {\bibfnamefont {T.}~\bibnamefont {Yamazaki}}, \ and\ \bibinfo {author}
  {\bibfnamefont {R.}~\bibnamefont {Kubo}},\ }\href
  {https://link.aps.org/doi/10.1103/PhysRevB.20.850} {\bibfield  {journal}
  {\bibinfo  {journal} {Phys. Rev. B}\ }\textbf {\bibinfo {volume} {20}},\
  \bibinfo {pages} {850} (\bibinfo {year} {1979})}\BibitemShut {NoStop}%
\bibitem [{\citenamefont {Goryo}\ \emph {et~al.}(2012)\citenamefont {Goryo},
  \citenamefont {Fischer},\ and\ \citenamefont {Sigrist}}]{Goryo2012}%
  \BibitemOpen
  \bibfield  {author} {\bibinfo {author} {\bibfnamefont {J.}~\bibnamefont
  {Goryo}}, \bibinfo {author} {\bibfnamefont {M.~H.}\ \bibnamefont {Fischer}},
  \ and\ \bibinfo {author} {\bibfnamefont {M.}~\bibnamefont {Sigrist}},\ }\href
  {https://link.aps.org/doi/10.1103/PhysRevB.86.100507} {\bibfield  {journal}
  {\bibinfo  {journal} {Phys. Rev. B}\ }\textbf {\bibinfo {volume} {86}},\
  \bibinfo {pages} {100507} (\bibinfo {year} {2012})}\BibitemShut {NoStop}%
\bibitem [{\citenamefont {Ryu}\ \emph {et~al.}(2010)\citenamefont {Ryu},
  \citenamefont {Schnyder}, \citenamefont {Furusaki},\ and\ \citenamefont
  {Ludwig}}]{ryu:2010}%
  \BibitemOpen
  \bibfield  {author} {\bibinfo {author} {\bibfnamefont {S.}~\bibnamefont
  {Ryu}}, \bibinfo {author} {\bibfnamefont {A.~P.}\ \bibnamefont {Schnyder}},
  \bibinfo {author} {\bibfnamefont {A.}~\bibnamefont {Furusaki}}, \ and\
  \bibinfo {author} {\bibfnamefont {A.~W.~W.}\ \bibnamefont {Ludwig}},\
  }\href@noop {} {\bibfield  {journal} {\bibinfo  {journal} {New Journal of
  Physics}\ }\textbf {\bibinfo {volume} {12}},\ \bibinfo {pages} {065010}
  (\bibinfo {year} {2010})}\BibitemShut {NoStop}%
\bibitem [{\citenamefont {Liu}\ \emph {et~al.}(2013)\citenamefont {Liu},
  \citenamefont {Shan}, \citenamefont {Yao}, \citenamefont {Yao},\ and\
  \citenamefont {Xiao}}]{Liu2013}%
  \BibitemOpen
  \bibfield  {author} {\bibinfo {author} {\bibfnamefont {G.-B.}\ \bibnamefont
  {Liu}}, \bibinfo {author} {\bibfnamefont {W.-Y.}\ \bibnamefont {Shan}},
  \bibinfo {author} {\bibfnamefont {Y.}~\bibnamefont {Yao}}, \bibinfo {author}
  {\bibfnamefont {W.}~\bibnamefont {Yao}}, \ and\ \bibinfo {author}
  {\bibfnamefont {D.}~\bibnamefont {Xiao}},\ }\href {\doibase
  10.1103/PhysRevB.88.085433} {\bibfield  {journal} {\bibinfo  {journal} {Phys.
  Rev. B}\ }\textbf {\bibinfo {volume} {88}},\ \bibinfo {pages} {085433}
  (\bibinfo {year} {2013})}\BibitemShut {NoStop}%
\bibitem [{\citenamefont {He}\ \emph {et~al.}(2018)\citenamefont {He},
  \citenamefont {Zhou}, \citenamefont {He}, \citenamefont {Yuan}, \citenamefont
  {Zhang},\ and\ \citenamefont {Law}}]{Law}%
  \BibitemOpen
  \bibfield  {author} {\bibinfo {author} {\bibfnamefont {W.-Y.}\ \bibnamefont
  {He}}, \bibinfo {author} {\bibfnamefont {B.~T.}\ \bibnamefont {Zhou}},
  \bibinfo {author} {\bibfnamefont {J.~J.}\ \bibnamefont {He}}, \bibinfo
  {author} {\bibfnamefont {N.~F.~Q.}\ \bibnamefont {Yuan}}, \bibinfo {author}
  {\bibfnamefont {T.}~\bibnamefont {Zhang}}, \ and\ \bibinfo {author}
  {\bibfnamefont {K.~T.}\ \bibnamefont {Law}},\ }\href {\doibase
  10.1038/s42005-018-0041-4} {\bibfield  {journal} {\bibinfo  {journal}
  {Communications Physics}\ }\textbf {\bibinfo {volume} {1}},\ \bibinfo {pages}
  {40} (\bibinfo {year} {2018})}\BibitemShut {NoStop}%
\bibitem [{\citenamefont {M\"ockli}\ and\ \citenamefont
  {Khodas}(2018)}]{Khodas}%
  \BibitemOpen
  \bibfield  {author} {\bibinfo {author} {\bibfnamefont {D.}~\bibnamefont
  {M\"ockli}}\ and\ \bibinfo {author} {\bibfnamefont {M.}~\bibnamefont
  {Khodas}},\ }\href {\doibase 10.1103/PhysRevB.98.144518} {\bibfield
  {journal} {\bibinfo  {journal} {Phys. Rev. B}\ }\textbf {\bibinfo {volume}
  {98}},\ \bibinfo {pages} {144518} (\bibinfo {year} {2018})}\BibitemShut
  {NoStop}%
\bibitem [{\citenamefont {Zinkl}(shed)}]{Zinkl}%
  \BibitemOpen
  \bibfield  {author} {\bibinfo {author} {\bibfnamefont {B.}~\bibnamefont
  {Zinkl}},\ }\href@noop {} {\bibfield  {journal} {\bibinfo  {journal} {et.
  al.}\ } (\bibinfo {year} {to be published})}\BibitemShut {NoStop}%
\bibitem [{\citenamefont {Brydon}\ \emph {et~al.}(2014)\citenamefont {Brydon},
  \citenamefont {Das~Sarma}, \citenamefont {Hui},\ and\ \citenamefont
  {Sau}}]{Brydon}%
  \BibitemOpen
  \bibfield  {author} {\bibinfo {author} {\bibfnamefont {P.~M.~R.}\
  \bibnamefont {Brydon}}, \bibinfo {author} {\bibfnamefont {S.}~\bibnamefont
  {Das~Sarma}}, \bibinfo {author} {\bibfnamefont {H.-Y.}\ \bibnamefont {Hui}},
  \ and\ \bibinfo {author} {\bibfnamefont {J.~D.}\ \bibnamefont {Sau}},\ }\href
  {\doibase 10.1103/PhysRevB.90.184512} {\bibfield  {journal} {\bibinfo
  {journal} {Phys. Rev. B}\ }\textbf {\bibinfo {volume} {90}},\ \bibinfo
  {pages} {184512} (\bibinfo {year} {2014})}\BibitemShut {NoStop}%
\bibitem [{\citenamefont {She}\ \emph {et~al.}(2017)\citenamefont {She},
  \citenamefont {Kim}, \citenamefont {Fennie}, \citenamefont {Lawler},\ and\
  \citenamefont {Kim}}]{She2017}%
  \BibitemOpen
  \bibfield  {author} {\bibinfo {author} {\bibfnamefont {J.-H.}\ \bibnamefont
  {She}}, \bibinfo {author} {\bibfnamefont {C.~H.}\ \bibnamefont {Kim}},
  \bibinfo {author} {\bibfnamefont {C.~J.}\ \bibnamefont {Fennie}}, \bibinfo
  {author} {\bibfnamefont {M.~J.}\ \bibnamefont {Lawler}}, \ and\ \bibinfo
  {author} {\bibfnamefont {E.-A.}\ \bibnamefont {Kim}},\ }\href {\doibase
  10.1038/s41535-017-0063-2} {\bibfield  {journal} {\bibinfo  {journal} {npj
  Quantum Materials}\ }\textbf {\bibinfo {volume} {2}},\ \bibinfo {pages} {64}
  (\bibinfo {year} {2017})}\BibitemShut {NoStop}%
\end{thebibliography}%

\end{document}